\documentclass[12pt,a4paper,final]{iopart}

\usepackage{breqn}
\usepackage{iopams}  
\usepackage{graphicx}
\usepackage[breaklinks=true,colorlinks=true,linkcolor=blue,urlcolor=blue,citecolor=blue]{hyperref}
\usepackage{bigfoot}
\usepackage{pdflscape}

\begin{document}

\title[State-by-state spectra fitting]{State-by-state emission spectra
  fitting for non-equilibrium plasmas: OH spectra of surface barrier discharge at argon/water interface}
  
  %{State-by-state emission spectra
 % fitting for non-equilibrium plasmas: application to barrier discharge emerging from water surface}

\author{Jan Vor\'a\v{c}, Petr Synek, Vojt\v{e}ch Proch\'azka, Tom\'a\v{s} Hoder}

\address{Department of Physical Electronics, Masaryk University,
	Kotl\'a\v{r}sk\'a 2, 61137 Brno, Czech Republic}

\ead{vorac@mail.muni.cz}

\begin{abstract}
Optical emission spectroscopy applied to non-equilibrium plasmas in molecular gases can give important information on basic plasma parameters, including the rotational, vibrational temperatures and densities of the investigated radiative states. 
In order to precisely understand the non-equilibrium of rotational-vibrational state distribution from investigated spectra without limiting presumptions, a state-by-state temperature-independent fitting procedure is the ideal approach. 
In this paper we present a novel software tool developed for this purpose, freely available for scientific community. The introduced tool offers a convenient way to construct Boltzmann plots even from partially overlapping spectra, in user-friendly environment.

We apply the novel software to the challenging case of OH spectra in
surface streamer discharges generated from the triple-line of
argon/water/dielectrics interface. After the barrier discharge is
characterised by ICCD and electrical measurements, the spatially and
phase resolved rotational temperatures from N$_2$(C-B) and OH(A-X)
spectra are measured, analysed and compared. The precise analysis shows that OH(A) states with quantum numbers $(v'=0,~9\leq N' \leq 13)$ are overpopulated with respect to the found two-Boltzmann distribution. We hypothesise that fast vibrational-energy transfer is responsible for this phenomenon observed here for the first time. Finally, the vibrational temperature of the plasma and the relative populations of hot and cold OH(A) states  are quantified spatially and phase resolved.

\end{abstract}

%52.30.-q 	Plasma dynamics and flow
%52.50.Dg 	Plasma sources
%52.70.-m 	Plasma diagnostic techniques and instrumentation
\vspace{2pc}

\submitto{\JPD}
%\maketitle

\newpage

\section{Introduction}
\label{s:intro}
Optical emission spectroscopy is a powerful non-invasive tool to
investigate non-equilibrium plasmas
\cite{ochkin2009spectroscopy,dilecce,simek,laux,brug}. Nowadays,
crucial parameters such as electric field \cite{sret,navratil,hoder}, 
electron density \cite{nikiforov,pu,mario} or densities of important 
species~\cite{vavsina2015determination} can be determined by
methods of optical emission spectroscopy with high resolution and
practically for any plasma of interest. In the case of the molecular
spectra, also the rotational temperature as a measure for gas
temperature can be obtained. Specair
programme~\cite{laux2002radiation} offers a commercial solution and is
nowadays widely used
for this purpose. Other programmes like
LIFBASE~\cite{luque1999lifbase} 
or PGOPHER~\cite{pgopher} are available for free and their functionality partially
overlaps with Specair.  
Nevertheless, due to the transient nature of plasmas and various
(de)excitation mechanisms participating in different timescales on
population of different rotational states, the determination of the
rotational temperature, or densities of these states, is not simple. A
typical example is the spectroscopy of the OH radical in water
containing discharges \cite{brug1}. 
%However, similar problematics can be found for other molecules as
%well. 
A more detailed approach is needed to investigate these spectra in
non-equilibrium plasmas as recently reviewed in \cite{brug2}. 

In this article, we present a new tool for state-by-state temperature-independent spectra fitting for molecular spectra of non-equilibrium
plasmas. We combine this approach with our previously published tool
{\it massiveOES}~\cite{massiveOES, vorac2017batch} for processing of large spectroscopic data (semi-automated methods become very handy if large amount of spatially or temporally highly resolved spectra has to be processed). We apply these tools to investigate the spatially and phase resolved OH spectra in surface barrier discharge emerging at the triple-line (triple-junction) \cite{buehrle} of argon/water/fused silica interface. Plasmas generated at the triple-line at atmospheric pressure feature usually high temperature and electric field gradients as a result of the presence of strong charge separation in the narrow sheath \cite{pitchford,stanfieldAPL}. This feature can be important in treatment of sensitive or bio-medical samples \cite{yusupov}, or when trying to understand the detailed plasma-initiated chemistry in surface treatment of polymers \cite{pavlinak}.

In general, the interest to study water-containing plasmas increased in last years enormously \cite{roadmap}. This fact is in direct connection to plasma applications in waste-water treatment \cite{locke02,thagard} and especially in plasma-medicine \cite{woedtke}. Due to the fact that these applications are operated in atmospheric pressure gases, the nature of these plasmas is highly transient and detailed and spatiotemporally resolved spectra analysis is required to understand their non-equilibrium. 
We hope that for this purpose the tool introduced in this article will be welcome by scientific community dealing with molecular spectroscopy for plasmas in various applications, not only at atmospheric pressure.

%In plasma jets, or barrier discharges under given conditions, the ionisation mechanism is a streamer, having typical time scales in nanoseconds. To find the the lead/connection between the chemistry happening in the active plasma and the resulting induced chemistry in the water or living tissue is important for precise plasma-source design [citeWinter]. The most important approaches typically used to study these plasmas are optical emission and laser-aided spectroscopy [reviewBrugemann,Dilecce,Simek]. 

%\vspace{0.5cm}
%
%In this paper ... 
%
%- state-by-state fitting approach
%
%- OH radical importance and its sensitivity to different mechanisms (advantage or disadvantage), exampled in his contribution
%
%- connection to long-term chemistry in water
%
%\vspace{0.5cm}
%
%As an example, we analyse here a surface discharge emerging at the so called triple-line [cite] created by argon gas/water interface at quartz barrier. ...

\section{Experimental setup and data acquisition}
\label{s:experimental}
\begin{figure}
\begin{center}
a)~
  \includegraphics[width = 0.45\textwidth]{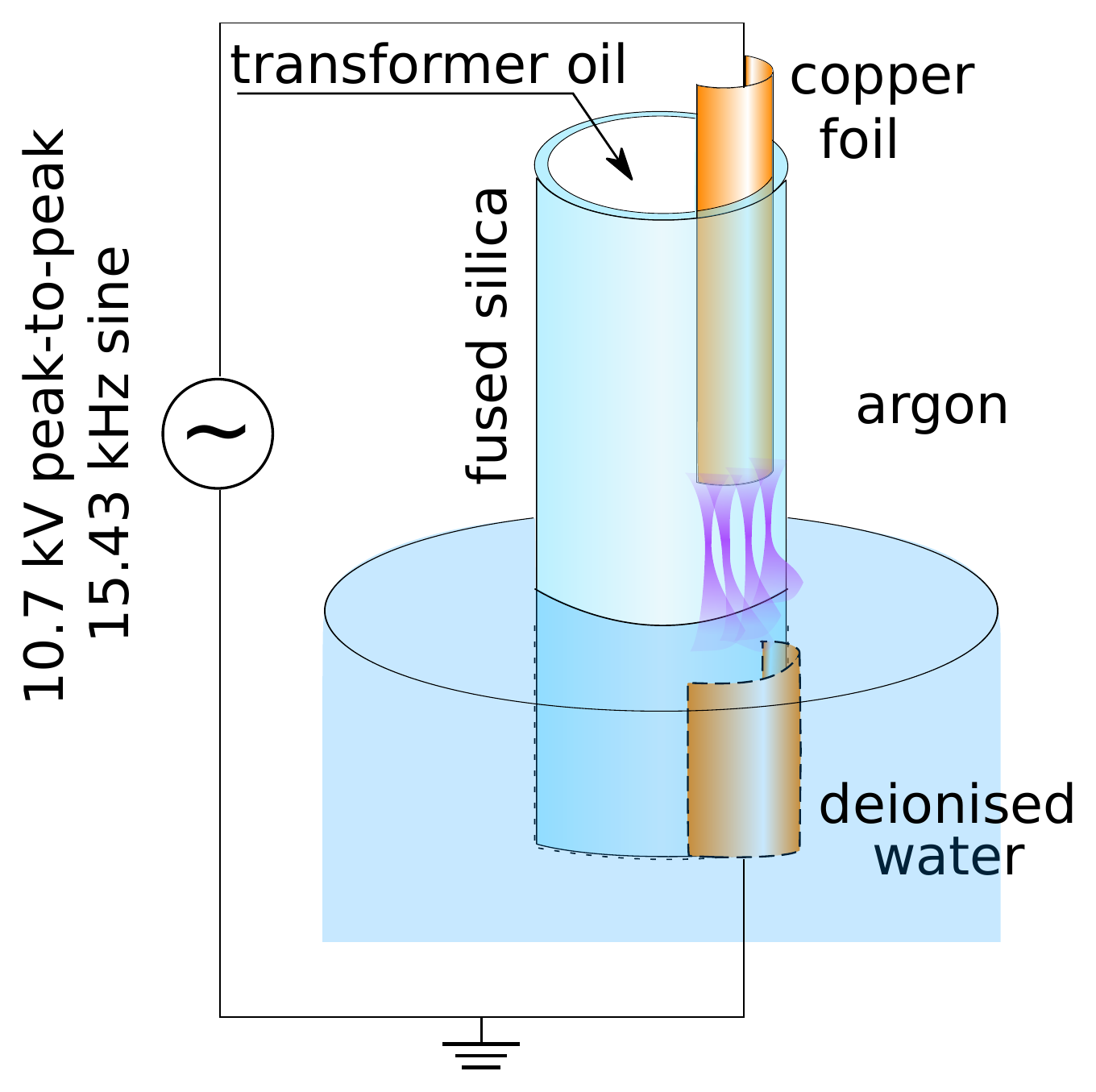}
  b)~
  \includegraphics[width = 0.45\textwidth]{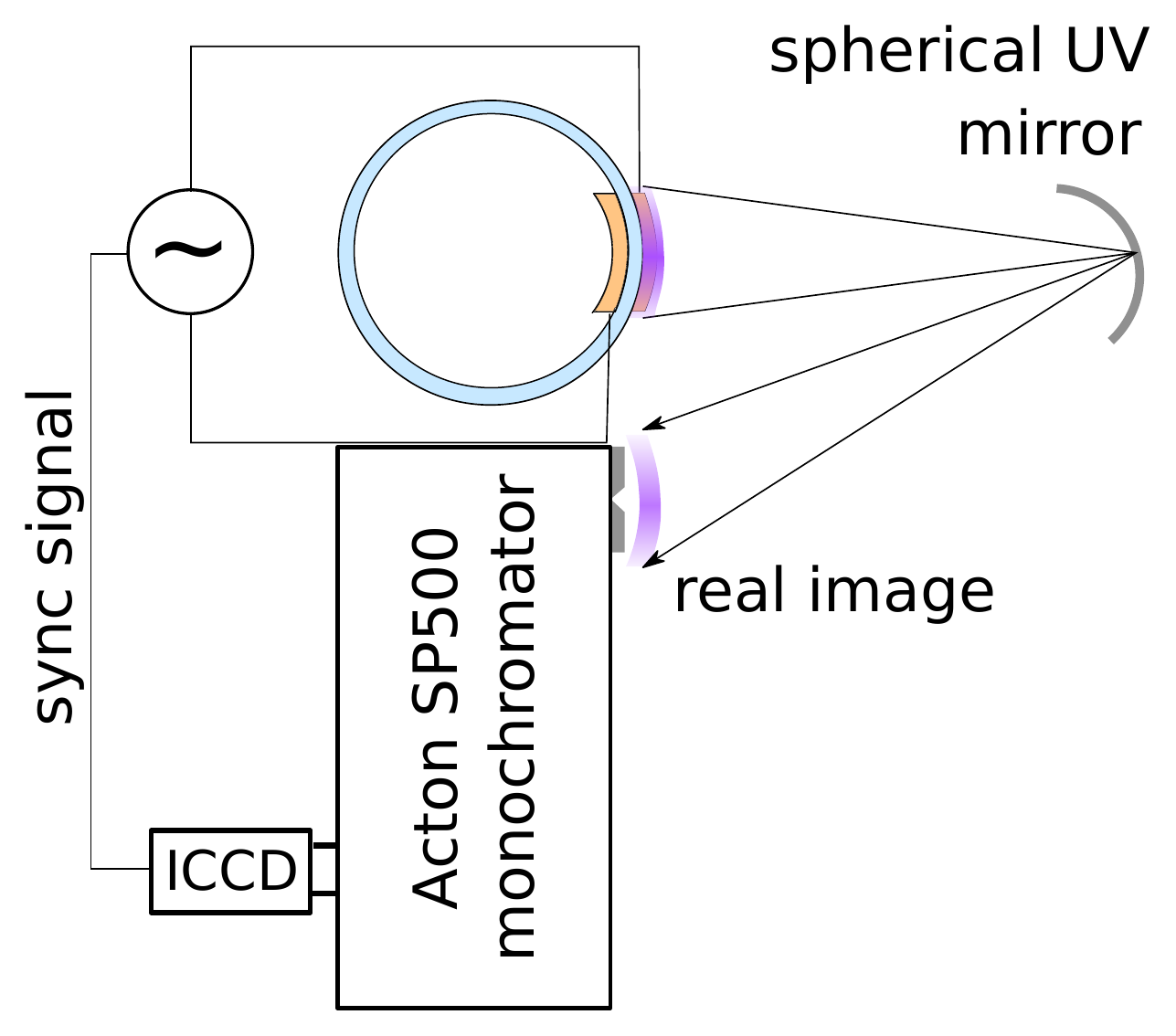}
  \end{center}
  \caption{Experimental setup: (a) Schematic sketch of the discharge
    reactor (detailed description of electrical circuit is omitted for
    brevity), (b) the scheme of optical detection setup.}
  \label{f:setup}
\end{figure}

The electrode arrangement is based on the standard surface barrier
discharge configuration, see figure~\ref{f:setup}. Two metal
electrodes (copper foil) are glued to opposite sides of a dielectric
sheet (fused silica). The dielectric sheet is curved and forms a
jacket of a cylindrical cuvette such that one electrode is glued on
the outside and the other is on the inside. The cuvette is filled with
transformer oil, so the discharges are forced to appear on the
outside. The cuvette with the electrodes is placed in a sealed chamber
rinsed with argon (1.4\,slm flow). The purity of the environment was checked indirectly by observing nitrogen
bands in the discharge emission. Based on the fact that nitrogen bands
intensity was at least two orders of magnitude smaller than the
bands of OH radical or argon lines, the residual nitrogen
content is considered negligible. In this particular experiment we add
1\,sccm of nitrogen gas to argon on purpose to get second molecular spectra to compare 
results of OH measurement with\footnote{Results for OH measurements show no
difference for discharge with and without the 1\,sccm of nitrogen gas flow.}. 

The chamber is partially flooded
with deionised water such that the outer electrode is submerged. In
this way, the water forms an additional dielectric barrier and the
discharges are thus not in direct contact with the electrode. 

Whole circuit consists of reactor chamber and parallel ballast capacitance 
to minimize resonance shifts due to surface charging (and consequently prevent fluctuation of applied voltage amplitude).
The outer submerged electrode is earthed, whereas the inner is driven by
sinusoidal alternating voltage with 5.4\,kV amplitude at resonant frequency
of 15\,kHz. 
Electrical measurements were performed with voltage probe (Tektronix
P6015A) and self assembled BNC current probe both connected to a
high-resolution and high-sampling rate oscilloscope (Keysight DSO-S
204A). Imaging of the discharge was performed with an ICCD camera (PI-MAX3 1024RB-25-FG-43) equipped with standard objective lens.

For spectroscopic measurements, a real image of the discharge is formed at an input slit of a
monochromator (Acton SP-500) using a spherical mirror (25.4\,mm
diameter, 100\,mm focal length, UV enhanced aluminium surface,
Thorlabs). The angle between the light direction and the optical axis
of the mirror is kept as small as possible to minimise the astigmatic
deformation of the image. By using a mirror we can avoid chromatic
error. This is critical particularly for argon discharges with water
vapour, as the emission is found particularly in the near UV and near
IR, spectrally separated by hundreds of nanometres, causing serious
shifts in the optical power of lenses. The light enters the
monochromator through a 20\,$\mu$m slit, is dispersed by a
grating (3600 grooves/mm, HUV blaze) and collected by an ICCD camera
(PI-MAX3 1024RB-25-FG-43). A thin line parallel to the electric field vector in
the centre of the electrodes was observed with radial width of the slit opening. 
This gives a radial selection of obtained spectra. The whole discharge length
of ca. 9\,mm was imaged to 780\,pixels, giving theoretical spatial
resolution of 11\,micrometers along the discharge axis. Due to weak signal, however, spatial
integration was necessary as part of the data processing, reducing the
spatial resolution to few tenths of a millimetre, see
section~\ref{s:processing} for details. 

The camera exposure was synchronised with the discharge driving
voltage. The positive and negative half-cycles were acquired
separately. To obtain sufficient signal-to-noise ratio, the signal was
accumulated over 10$^6$ discharge periods on the CCD chip before
readout. The (0,0), (1,1) and (2,2) bands of OH (A\,$^2\Sigma^+
\rightarrow \mathrm X\,^2\Pi$) transition in the spectral region from
306\,nm to 318\,nm were observed.
Because the grating dispersed too well, it was not possible
to capture the spectral region of interest at once. Instead, three
adjacent spectral windows were acquired, in the respective ranges
(305.84--310.86)\,nm, (308.91--313.89)\,nm and
(312.82--317.75)\,nm. The last spectral window was acquired with
$3\times10^6$ accumulations to enhance the signal-to-noise ratio of
the weaker spectral lines with higher $J'$ and $v'$ quantum numbers.

Another spectral window in the range (333.67--338.61)\,nm was
recorded for reference. This window contains the (0,0) band of the N$_2$
(C\,$^3\Pi_{\mathrm u} \rightarrow \mathrm B\,^3\Pi_{\mathrm g}$)
transition. The rotational temperature obtained from this band is
believed to reliably reflect the translational temperature of
gas~\cite{vorac2017batch, hsieh2016analysis, cech2009influence, bruggeman2012absolute}.

In further text, the electronic states will be labelled only by the first
letter for brevity, e.g. state N$_2$(C) or transition N$_2$(C-B).

\section{Data processing}
\label{s:processing}
\begin{figure}
  \includegraphics[width = \textwidth]{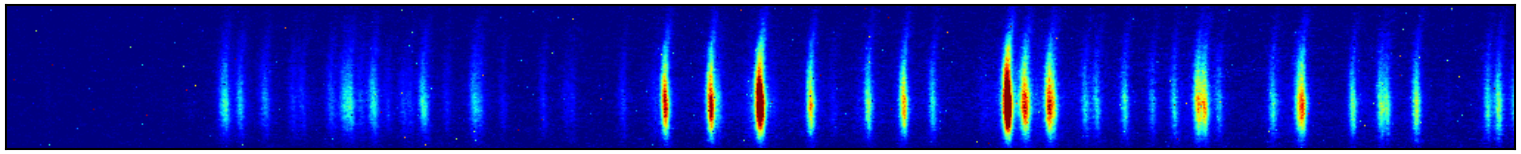}
  \includegraphics[width = \textwidth]{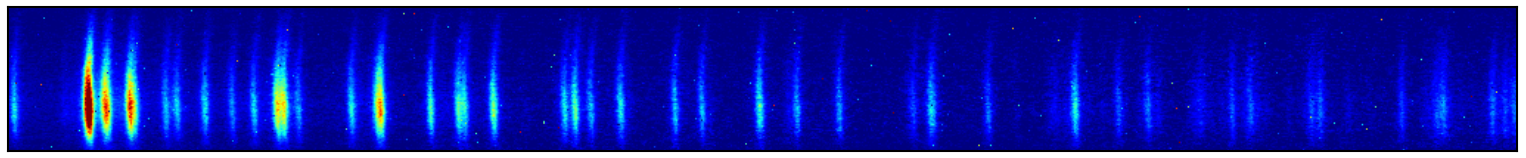}
  \includegraphics[width = \textwidth]{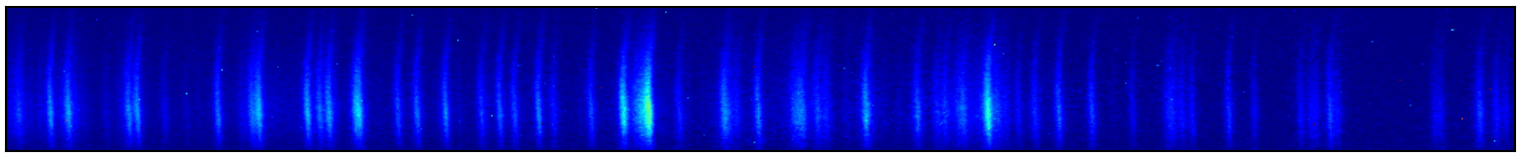}
  \includegraphics[width = \textwidth]{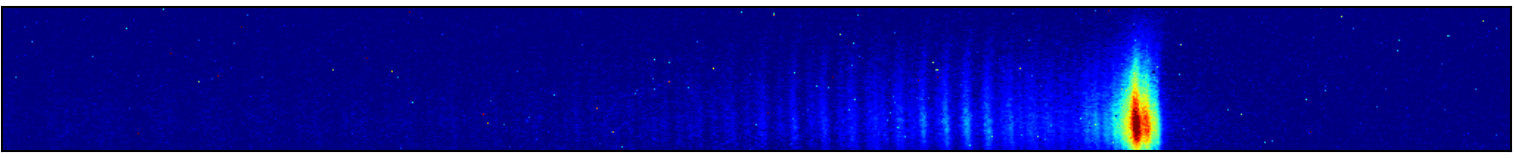}
  \caption{An example of spectroscopic image of the discharge in the 
    positive half-cycle of the applied voltage after dark current subtraction and
    spectral sensitivity correction. Spectral ranges
    (305.84--310.86)\,nm, (308.91--313.89)\,nm, (312.82--317.75)\,nm
    and (333.67--338.61) are shown, respectively.  The shown images
    are vertically cropped to the region of interest and partially
    vertically binned with a bin size of 7 pixels. The vertical
    distance corresponds to axial position. The triple-line interface
    is at the bottom and the discharges propagate upwards. }
  \label{f:4frames}
\end{figure}

The dark current was subtracted from the acquired images. The relative
spectral sensitivity was determined using a calibrated deuterium lamp
and the images were corrected accordingly. Examples
are shown in figure~\ref{f:4frames}. The shown examples are vertically
cropped to the region of interest and vertically binned with a bin
size of 7 pixels to emphasise an unwanted deformation of the image
in the monochromator. The input slit is imaged as a curved line, the
wavelength calibration must therefore be performed for each vertical position
accordingly. In this work, we assume, that the deformation results in
constant shift of the spectrum along the wavelength-axis and the
wavelength resolution is constant. The wavelength calibration is
performed by comparing the measured spectrum to a simulation, as
described below.

\begin{figure}
  \includegraphics[width = \textwidth]{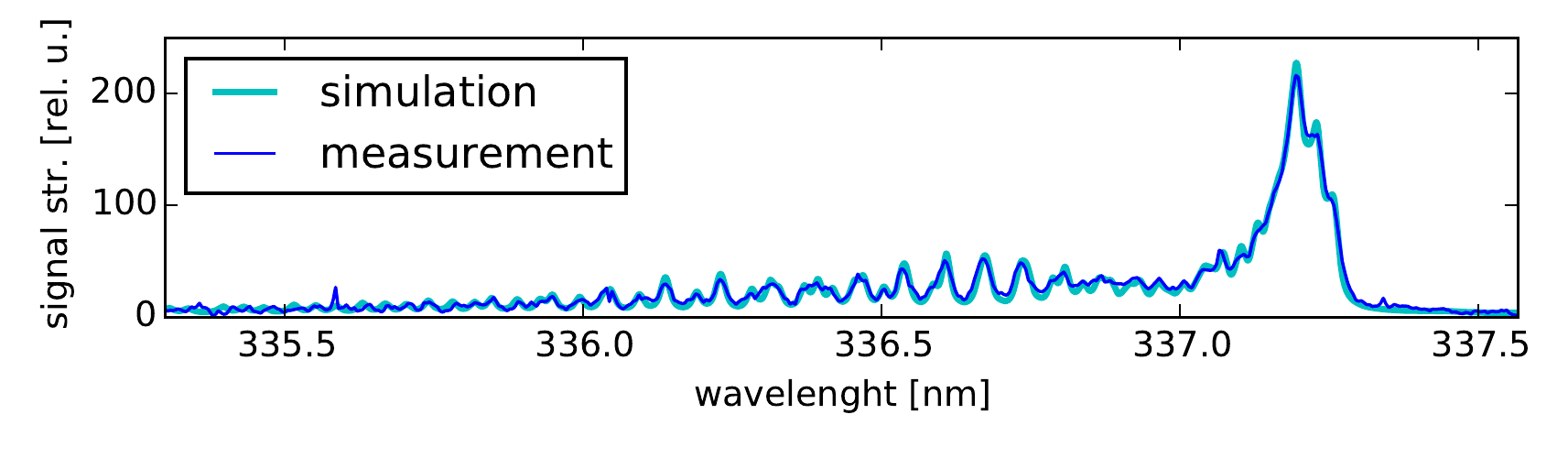}
  \caption[]{An example of the N$_2$(C-B) (0,0) band emitted by positive
    streamer discharges at water interface. The shown spectrum was measured at the
    point of the triple-line (bottom part of the discharge). The
    shown simulation
      is for rotational temperature of 371.7\,K.}
  \label{f:N2fit}
\end{figure}

The rotational temperature of N$_2$(C) is obtained by comparing the
measurement with a simulation\footnote{Please note that all shown simulations
      have y-data proportional to photon flux, {\it not} 
      intensity.}, minimising the sum of squared
residuals. Details are described in~\cite{vorac2017batch}. First, the
wavelengths of the measurement and of the simulation must be
matched. For this step, a synthetic spectrum with roughly correct
temperature is prepared. Corresponding positions in the simulation and
the measurement are identified by visual comparison. This is tricky
particularly for the (0,0) band of N$_2$(C-B) transition, as only one
band head is available at 337.2\,nm and the maximum of the R-branch
around 336.7\,nm shifts with rotational temperature. Other $(v', v''=v')$ 
bands of
N$_2$(C-B) are too weak due to significantly smaller Franck-Condon
factors~\cite{laux1992arrays} and are thus unobservable even at high
vibrational temperatures~\cite{vorac2017batch}. 
In more energetic discharges with
hydrogen content, also emission of NH(A\,$^3\Pi \rightarrow \mathrm
X\,^3\Sigma^-$) (0,0) band at 336\,nm can be
observed~\cite{vorac2017batch, janda2016escampig}, interfering with
the N$_2$ emission, but sometimes enabling more reliable wavelength
calibration. In our case, however, the NH(A-X) emission was not
observed. With the 3600\,g/mm grating, also finer features in the band
head were observable, which helped to match the wavelengths of the
simulation and measurement correctly. Nevertheless, an iterative
approach, of adjusting the wavelength calibration and the rotational
temperature until the best possible fit is obtained, is strongly
advised. Wrong wavelength calibration results in serious error of the
rotational temperature. Although non-equilibrium population 
distribution of rotational levels
of the N$_2$(C) state were reported
e.g. in~chapter 4.2.4 of \cite{ochkin2009spectroscopy},
it was not observed in our conditions. Indeed, the theoretical
spectrum, assuming Boltzmann distribution, fits the experimental data
very well, see figure~\ref{f:N2fit}.
After the three coefficients\footnote{The wavelength of a measured 
point is calculated
  from pixel position by second order polynomial.}
 for wavelength calibration were
found, the constant
offset term was optimised for each row of the measurement. In this
step, the partially binned measurements shown in
figure~\ref{f:4frames}  for wavelength calibration were used for the variable wavelength
calibration. The signal strength was insufficient for eventual
corrections of the first and second order coefficients, these were
thus considered constant. Closer inspection of all fits did not reveal
any serious violation of this assumption. 

When all wavelength calibration coefficients were known, the spatial
integration was performed. To avoid unnecessary losses of spectral
resolution, a spectra-matching function from 
{\it massiveOES}\footnote{The source code is available at
  the reference URL. The mentioned function is \verb|match_spectra()|
from the file \verb|spectrum.py|.} package
was used~\cite{massiveOES, vorac2017batch}. For reliable spectral fitting, sufficient
signal-to-noise ratio is necessary. This can be enhanced by spatially
integrating over wider range. Spatial integration, on the other hand,
reduces spatial resolution. To achieve the optimal trade-off, a
threshold for signal strength (fivefold of the maximal strength in the
ensemble of partially binned data) was selected. The limits for the
spatial integration were then selected such that each spatial range
after integration gave sufficient signal. In this way, the spatial
resolution is variable but never reduced more than necessary. 

This approach resulted in two \verb|massiveOES.MeasuredSpectra()|
instances, containing spatially resolved spectra of N$_2$(C-B) for
positive and negative streamer discharges (see
section~\ref{s:electro} for explanation of the labels). These were
then processed in the massiveOES programme assuming Boltzmann
distribution to obtain the rotational temperatures as in~\cite{vorac2017batch}.

The pre-processing of the OH spectra was analogous. The
wavelength-matching had to be performed for each spectral window
separately.
The identification of
OH(A-X) spectra is more satisfying than that of N$_2$(C-B). The whole
matter was, however, complicated by the fact, that in our case, the
spectrum did not follow a single-temperature Boltzmann
distribution. For the sake of matching the measurement with the
simulation, it is advisable\footnote{This applies to discharges in
  contact with liquid water~\cite{bruggeman2009rotational,
    nikiforov2011physical} 
and with excessive
  abundance of water vapour. There are also
  discharges with equilibriated distribution of OH(A) 
rotatinal levels~\cite{vorac2017batch}.} to assume 
lower temperature (hundreds K)
for wavelengths in the range (308--310)\,nm but higher temperature
(thousands K) for wavelengths above 310\,nm 
 (comp. results -- figure~\ref{f:OH_boltz}). The spatial integration
was also performed analogously, focusing on comparable signal strength
rather than on equal integration domains. 
 After that, the spectral windows were attached to a single
window. 
%This was then divided to spectra, analogously, 
%according to the spatial
%coordinate. 
These results were then compared with the simulation. 
Nevertheless, the fits
assuming a single-temperature Boltzmann distribution were
unsatisfactory, see example in figure~\ref{f:OH_fit}(a). A novel method for spectral evaluation was thus
developed and is described below. 

When the assumption of Boltzmann distribution fails, another approach
needs to be taken. The whole procedure of preparing a synthetic
spectrum\footnote{Using a database with wavelength--emission
  coefficient pairs, calculating line broadening and spectral matching to
  the measurement, see~\cite{vorac2017batch}.} is preserved, except
for assuming anything about the relative populations. Instead, a
single synthetic spectrum is calculated for {\it each rovibronic
  state} in question, i.e., state that emits lines in the observed wavelength
region. Relative population of each state is then treated as a fit
parameter. The resulting best-fitting spectrum is a linear combination of
these fractional spectra. 
The problem\footnote{Implementation can be found in the file
  \verb|MeasuredSpectra.py|, method \verb|fit_nnls()| of the class
  \verb|MeasuredSpectra|.} 
can be mathematically formulated as
\begin{equation}
\min \limits _{n_{(J', N', v')}} \left \|m(\lambda) - \sum \limits
_{(J', N', v')}
n_{(J', N', v')}\,s_{(J', n', v')}(\lambda) \right\|^2 
\end{equation}
with the restriction that  $n_{(J', N', v')} \geq 0$. In this formula,
$m(\lambda)$ is the measured spectrum as a function of wavelength
$\lambda$ and $s_{(J', N', v')}(\lambda)$ are the respective fractional
simulated spectra
for each upper state, defined in the same points as
$m(\lambda)$. $(J', N', v')$ are the three quantum numbers that define
a rovibronic state, for details see~\cite{herzberg, Vorac2014thesis}. 
The optimised parameters $n_{(J', N', v')}$ are the relative populations of the respective
rovibronic states. Example of a state-by-state fit is shown in figure~\ref{f:OH_fit}(b).
Dividing each with its degeneracy $(2\,J' + 1)$ and
plotting its logarithm versus the energy of the levels, a traditional Boltzmann
plot can be constructed and analysed. 

The state-by-state fitting functionality was incorporated into the
{\it massiveOES} package~\cite{massiveOES, vorac2017batch}, together with a convenient
tool to visualise contributions of selected states to the
spectrum. This facilitates the inspection of the fit and the
reliability can be judged easily. 

\section{Results and discussion}
\label{s:results}

In following chapters, we firstly describe the overall behaviour and features of the generated plasma. Using the intensified CCD imaging we reveal the spatial structure of the typical filaments for both polarities and generally describe their mechanisms. As a next step the undertaken electrical measurements clarify the temporal scale of the discharge current pulses and estimate the power balance during the applied voltage period. Finally, the spectroscopic analysis will be conducted on recorded spatially and phase-resolved spectra of already introduced discharge and the results of novel state-by-state temperature-independent fitting procedure will be discussed.

\begin{figure}
\begin{center}
a)~
  \includegraphics[width = 0.4\textwidth]{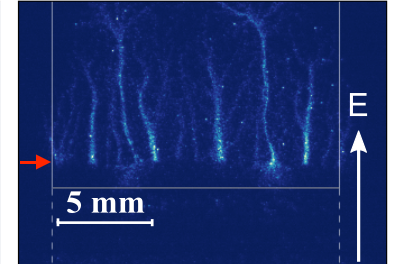}
b)~  
  \includegraphics[width = 0.4\textwidth]{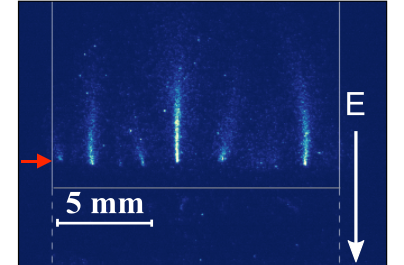}
  \end{center}
  \caption{Intensified CCD camera photographs of positive a) and negative streamer discharges b). The vertical grey lines denote the edges of cuvette and the horizontal the water surface. The red arrow denotes the elevated water interface wetting the fused silica surface of the cuvette, i.e. the triple-line. While for positive streamers the length and tree-like structure is similar for all filaments, for the negative streamer discharges the size and the intensity differ from case to case. At some cases just a cathode spot is visible, without a pronounced diffuse jet.}
  \label{foto}
\end{figure}

\subsection{Intensified CCD imaging and electrical characterization}
\label{s:electro}

After switching-on the applied voltage, the discharge starts in
irregular manner and it takes approximately ten minutes until the
spatiotemporally stable argon/water interface is established on the
surface of the fused silica dielectrics. In such stable mode the discharge
takes a filamentary form of usually several filaments propagating from
the approx. 1.5\,cm long triple-line upwards, along the dielectric
surface. This situation is depicted in Fig.\,\ref{foto} where
intensified CCD images of the discharge are taken for positive and
negative polarity of applied voltage. The discharges stochastically
appear at different positions, so at longer exposures, the plasma
appears spatially rather homogeneous. 

Typically, the surface barrier discharge polarity is designed by the voltage applied on to the upper (not embedded) electrode \cite{gib,gibalov}. We follow this notation although the voltage is applied to the embedded in oil isolated electrode (for safety reasons). As a result we describe the discharges propagating during the positive applied half-period as negative streamer discharges (NSD, propagating along the anodic dielectric from cathodic water interface) and positive streamer discharges (PSD) during the negative half-period of the applied voltage. 

The structure of the PSD shown in Fig.\,\ref{foto}a) is typical for surface positive streamers, resembling a pronounced tree-like structure with relatively homogeneous light emission intensity along its span. Taking into account the length of the filaments, the intensity of light emission and the time scale of the generated current pulses (see further in the text), we can say that the discharge mechanism of PSD is of streamer nature and no significant transition to the leader discharge (as reported e.g. in \cite{akishev}) takes places in our case. 

For NSD, the diffuse structure resembles a cometary shape with most intense light emission at the connection with the negative triple-line interface. These intensive spots at the negative electrode are sometimes called tufts \cite{stanfieldAIAA2014} or hot-spots of the strong sheath region (due to the elevated gas temperature at their location, see in \cite{stanfieldAPL}). We prefer the term cathode spots \cite{akishev,hoderPRE,stanfieldDIZ} as the high-electric-field region was identified as cathode layer created after the micro-sized positive streamer-like ionizing wave arrives to the cathode \cite{hoderPRE,grosch}. We assume that the same mechanism takes place also here at the negative triple-line water interface of the surface barrier discharge (compare \cite{grosch}).

\begin{figure}
\begin{center}
	\includegraphics[width = 0.35\textwidth]{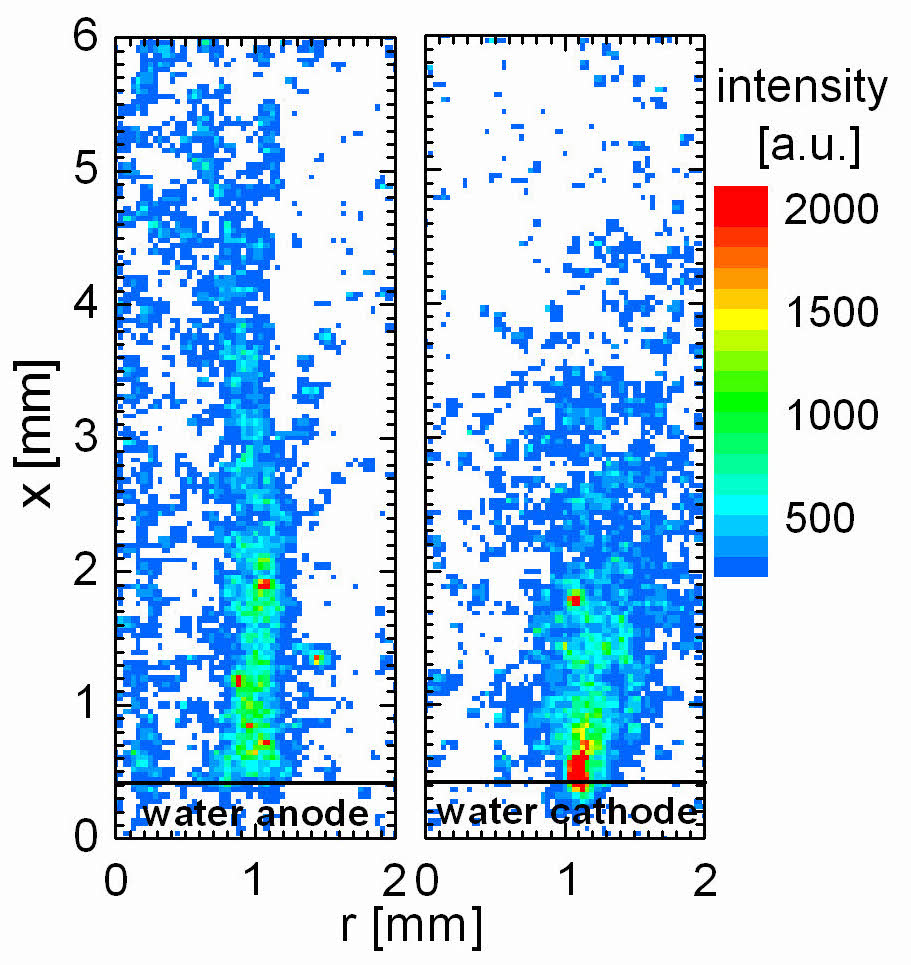}
	\end{center}
  \caption{Close-up ICCD images of the typical PSD (left) and NSD (right) discharge filaments emerging from the triple-line of deionised water interface and propagating along the fused silica dielectrics. }
  \label{interface}
\end{figure}

Analysing the single filament emission intensity in more details we can distinguish the above mentioned features in Fig.\,\ref{interface}. We picked up typical filaments for both polarities and plotted their emission intensity along their axial and radial dimensions. The elevated intensity of the cathode spots for the NSD at the triple-line is clearly visible. PSD resembles a tree with homogeneously intensive stem emission and branching while NSD is created by cathode spot and diffusive cloud above.

%
%After switching on the power the discharge starts in irregular manner
%where due to counterplay of electro-wetting \textcolor{red}{A je to
%  urcite electro-wetting? Neni to spis docasna modifikace povrchu
%  kyvety chemickou cestou, ktera vede k vyssi smacivosti?} and evaporation the individual discharges travel up and down on the cuvette. This mode last about 5-15 minutes after which balance of the water level on the cuvette surface reaches equilibrium and the discharge stabilizes. Stabilization can be observed both visually and on electrical measurements. All further discussion in this article regards to stable operation mode of the discharge.
%
%Camera imaging (see figure\,\ref{foto}) revealed that in both half cycles the discharge have strongest emission in the area close to the water meniscus. However discharge shape differs for each half period the the cycle. Discharges in half cycle with positive applied voltage resemble upside-down raindrop shape characteristic for negative streamer discharges while in opposite half cycle discharge have curvy tree-like shape which suggest positive streamer discharges.

\begin{figure}
\begin{center}
	\includegraphics[width = 0.5\textwidth]{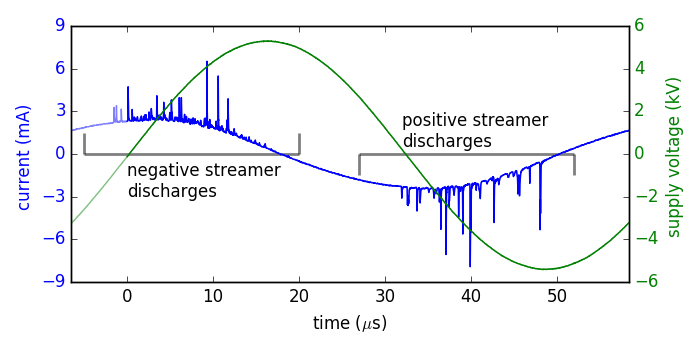}
	\end{center}
  \caption{Electrical measurement of current and voltage with denoted 'positive streamer discharges' and 'negative streamer discharges' intervals.}
  \label{el_over}
\end{figure}

The results of electrical measurements can be seen in Fig.\,\ref{el_over} 
showing the development of applied voltage and external circuit current waveforms.
Also, intervals of temporal integration of intensified CCD during which the imaging and spectral data were accumulated are denoted there. 
Note that first discharges for given voltage half-period starts during the end of the preceding one. This is so-called backward-discharging effect caused by the effectively high electric field in the gap advantageously biased by the residual surface charges from discharges of previous polarity \cite{gib}. 

\begin{figure}
\begin{center}
a)~
  \includegraphics[width = 0.45\textwidth]{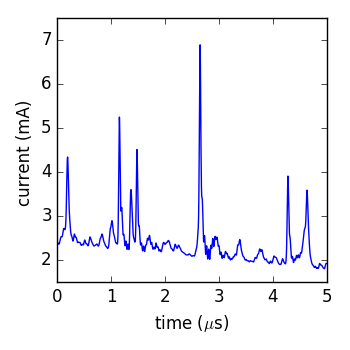}
  b)~
  \includegraphics[width = 0.45\textwidth]{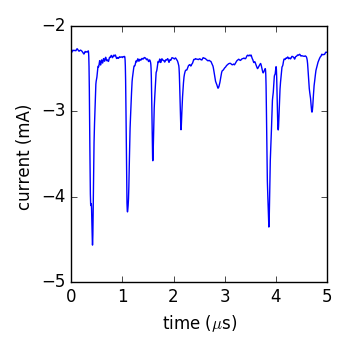}
  \end{center}
  \caption{More detailed views on current measurements for 'negative streamer discharges' (a) and 'positive streamer discharges' (b) intervals. Note large number of smallish current pulses in the 'negative streamer discharges' phase.}
  \label{el_detail}
\end{figure}

Close-ups of current pulses for each half period are shown in
figure\,\ref{el_detail}. Negative streamer discharges show pulses with mean full width half maximum (FWHM) of 15\,ns and with wide variety of amplitudes, from units down to tenths of milliamperes. Presumably, the creation process of some of the cathode spots is not finished and they are not able to feed the filament with sufficient current for given local conditions at the interface (locally stored surface charge, solvated electrons in water or generally change in conductivity of plasma-treated deionised water \cite{go}). Thus, their progress does not continue regularly to negative streamer which causes pulses of different amplitudes.
%that this points out to large number of electron avalanches spreading
%from the meniscus towards the positively charged electrode. 
During the negative
applied voltage half cycle, a lower number of PSD breakdowns occurs and they feature lower variation of amplitude and mean pulse FWHM of 30\,ns.
In both polarities the maximum peak values of the current pulses are few milliamperes, which is in agreement with measurements performed on barrier discharge filaments in atmospheric pressure argon \cite{kloc}.

Power loss in reactor chamber calculated from electrical measurements data is 1.45\,W. Rough estimate of power consumed in discharges is 0.8\,W while the rest goes to capacitive losses.

%\textcolor{green}{ ...to add rough estimation of the temperature change from this dissipated power PETRE}

%According to this, we have divided the the spectral measurements to two parts denoted as  \textbf{'positive streamer discharges'} and \textbf{'negative streamer discharges'}.

\subsection{Rotational temperature of N$_2$(C)}
\begin{figure}
\begin{center}
  \includegraphics[width = 0.45\textwidth]{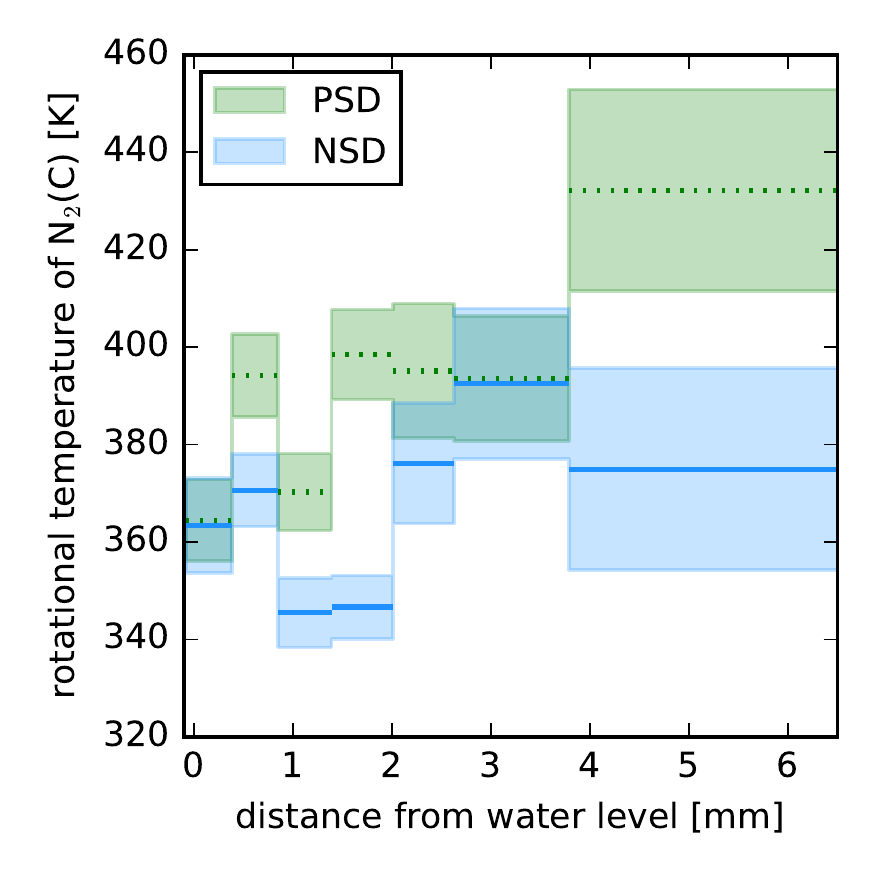}
  \end{center}
  \caption{Rotational temperature of N$_2$(C) along the
    discharge. PSD (dotted) -- positive streamer discharges, NSD (solid) --
    negative streamer discharges.}
  \label{f:N2T}
\end{figure}

The rotational temperatures of N$_2$(C) obtained from the fits are
shown in figure~\ref{f:N2T}. The spatial integration limits are
represented by the spans of the horizontal lines. The vertical span of
the coloured region illustrates an estimate of the 68\,\% confidence
interval (one standard deviation). This uncertainty is estimated from the fitting procedure itself,
assuming that the wavelength calibration and the instrumental
broadening are precisely known. It is therefore likely that the shown
uncertainty is underestimated. In figure~\ref{f:N2T} it can be seen that the gas
temperature is about 100 degrees higher than the room temperature
(ca. 294\,K). The power dissipated in the discharges is rather low (see
section~\ref{s:electro}), yet sufficient to significantly increase the temperature
in the discharge channels. 
Our measured temperatures are, as expected, lower than in the case of similar high-power industrially-used setup investigated in \cite{galmiz}.

The rotational temperature at the water level
(0--0.4\,mm) 
is remarkably similar for positive and negative streamer
discharges. In the case of positive streamers, the temperature rises
with the distance until the last observable point around 5\,mm. 
The rotational temperature near the triple-line of surface barrier
discharge is usually higher for positive streamers \cite{stanfieldDIZ}
with the exception of the location of cathode spots in opposite
polarity. The tendency of increasing rotational temperature with
increasing distance from the triple-line is unexpected for PSD, as it
is opposite to the surface barrier discharge case in air
\cite{stanfieldDIZ}. This can be caused by different heating
mechanisms of the positive streamer in argon than in air. Possibly, a
not fully developed leader initiation could affect the spectra, as the
transition to the leader mechanism in atmospheric pressure argon seems
to be much faster/easier than for air \cite{akishev}. In NSD however,
the concave tendency of the temperature dependency on spatial
coordinate is similar to that one observed in air
\cite{stanfieldDIZ}. The elevated temperature at the triple-line
interface is caused by the presence of cathode spots. There, in
high-gradient temperature and electric field region, the rotational
temperature determined from N$_2$(C) can under certain conditions even reach much higher values than in the positive streamer discharge. In order to resolve the exact mechanism of N$_2$(C) rotational energy distribution and heat transfer under given conditions, a well spatiotemporally resolved investigation is needed.

%
%\textcolor{red}{Prodiskutovat s nekym, kdo tomu rozumi. Muze pozitivni
%streamer nabirat cestou na energii? Neni to spis emise toho
%pseudo-doutnaveho vyboje? Proc by mela teplota rust se vzdalenosti?}
%
%\textcolor{red}{Co zaporne vyboje? Da se nejak vysvetlit ten pokles
%  teploty na 0.8\,mm? Proc to na hladine vyslo stejne a pak uz ne? A
%  je to vubec realny trend nebo se to topi v sumu, jenom tady
%  podcenujeme chyby?. Kazdopadne, z OH to vychazi dost podobne, akorat
%jsou vsechny trendy slabsi nez odhad chyb (vsechno se deje v ramci chyby).}

%{It also appears that the temperature in the
%positive streamer discharges is about 20\,K higher in the range
%1--2\,mm. This can be explained by the fact that in the positive
%streamer discharges, the local electric field is usually notably
%higher [\textcolor{red}{Tome, citace?}] and the positive streamers are generally more
%energetic than the negative ones [\textcolor{red}{jeste jedna
%    citace?}].  }

\subsection{OH(A-X) emission analysis}

%\begin{landscape}
\begin{figure}
\begin{center}
  \includegraphics[width = 0.9\textwidth]{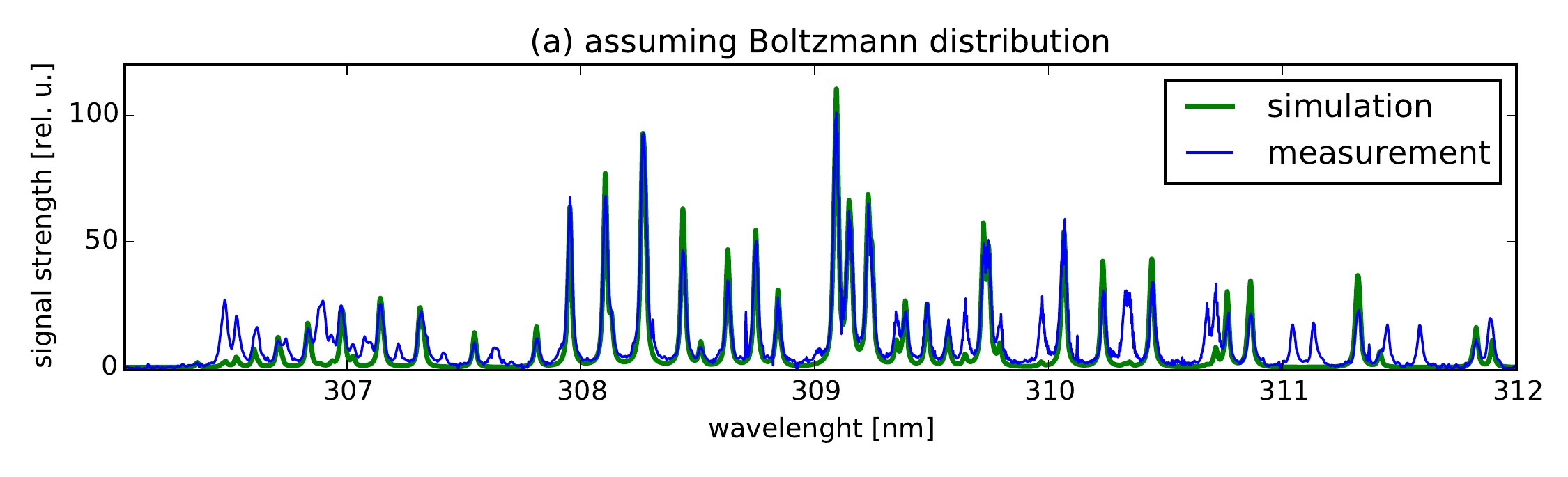}\\
  \includegraphics[width = 0.9\textwidth]{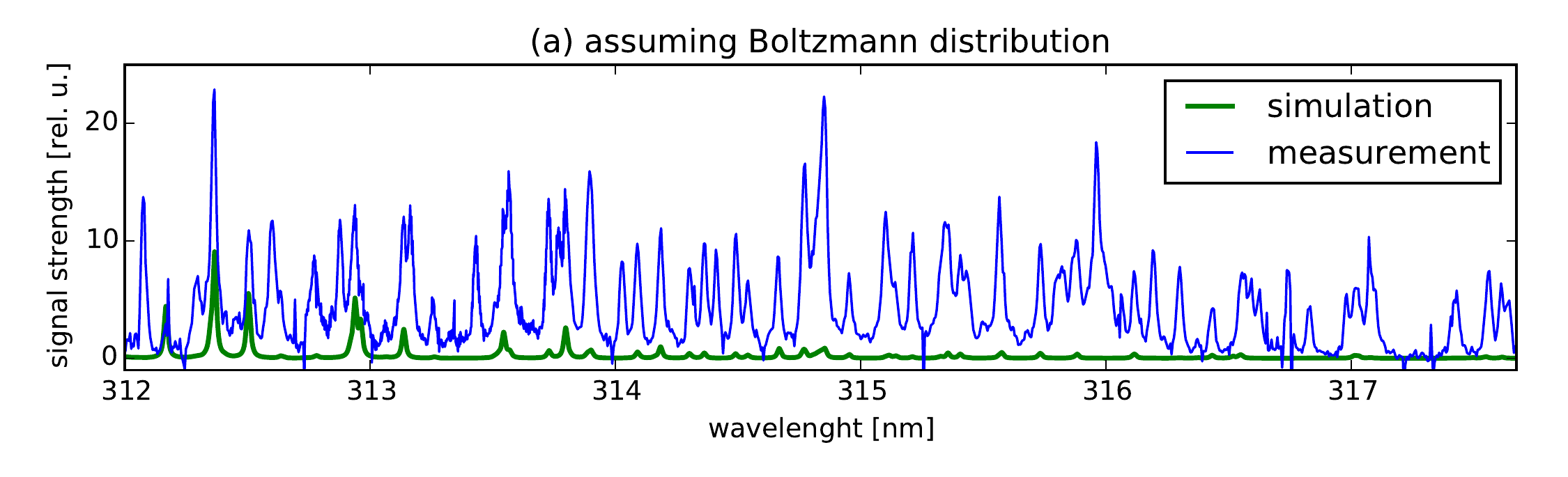}\\
  \includegraphics[width = 0.9\textwidth]{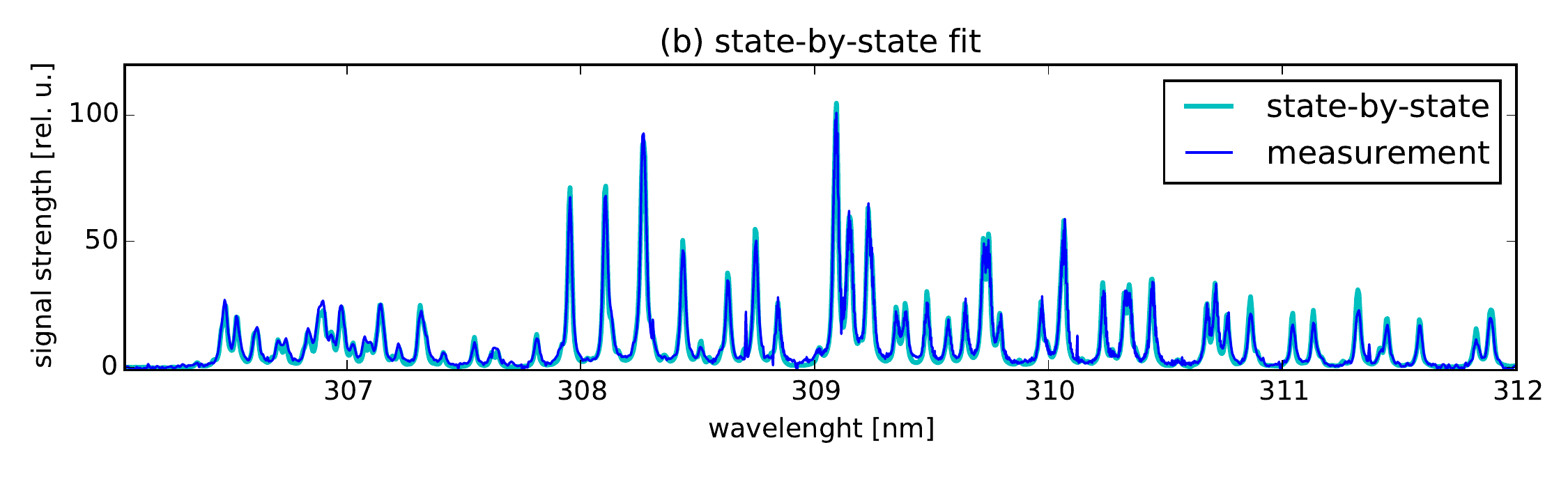}\\
  \includegraphics[width = 0.9\textwidth]{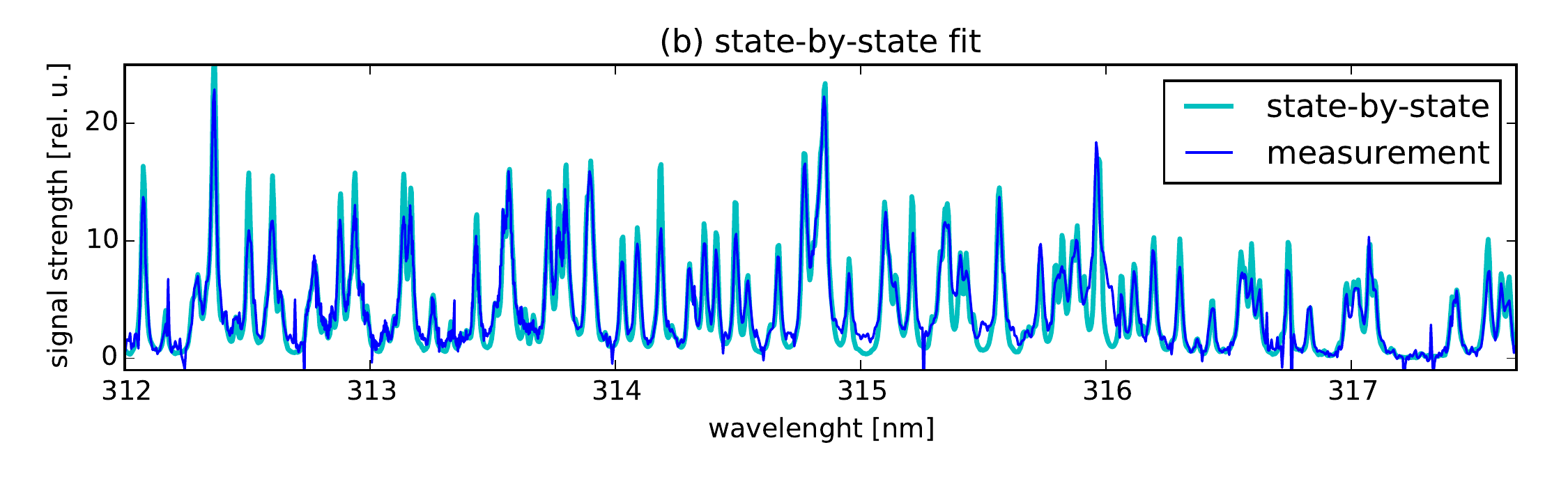}
  \end{center}
  \caption[]{Example of the fits (a) assuming Boltzmann distribution,
    best fitting $T_{\rm rot} = 848$\,K and $T_{vib} = 5\,300$\,K, 
    (b) state-by-state fit
    not assuming any population distribution. The Boltzmann plot
    corresponding to (b) is in figure~\ref{f:OH_boltz}.}
  \label{f:OH_fit}
\end{figure}
%\end{landscape}

\begin{figure}
\begin{center}
  \includegraphics[width = 0.9\textwidth]{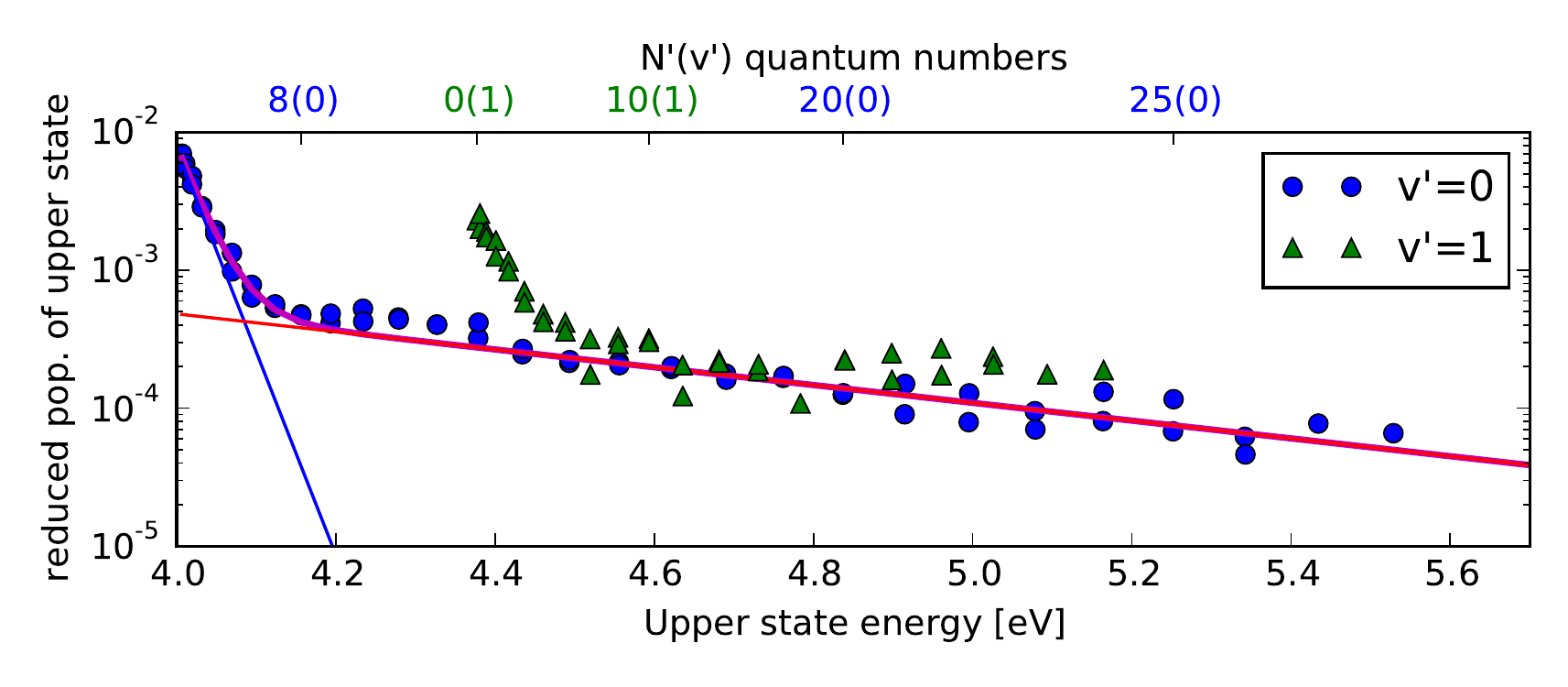}
  \end{center}
  \caption[]{Boltzmann plot from spectrum in
    figure~\ref{f:OH_fit}(b). The relative populations are divided by
    the respective degeneracy, i.e. {\it reduced}.
    The magenta curved line is a two-temperature fit, where the blue
    and red straight lines are its components.
    The two fitted rotational temperatures
    are $T_{\rm rot} ^{\rm low} = 343 \pm 28$\,K and 
    $T_{\rm rot} ^{\rm high} = 7\,800 \pm 550$\,K.
    Fitted populations of $v'=2$ states are not shown.}
  \label{f:OH_boltz}. 
\end{figure}

Boltzmann plots were constructed for each spatial coordinate from state-by-state spectra fitting.
In Fig.\,\ref{f:OH_fit}, an example of the fit is shown together with temperature-bound fit  for comparison. 
A typical example of Boltzmann plot is shown in figure~\ref{f:OH_boltz}. This result at
first sight resembles similar results obtained in works of 
others~\cite{ochkin2009spectroscopy,bruggeman2009rotational,nikiforov2011physical}.
It should be noted, that in this case the Boltzmann plot was obtained
using not a single rotational line, but using all lines from the
respective upper state at once, even the overlapping ones. The general
characteristics of the Boltzmann plot is the same as in the
references. 
We observe a
{\it cold group} and a {\it hot group} with greatly different
temperatures. 
We adapt the explanation of
Ochkin~\cite{ochkin2009spectroscopy}, that the cold group consists of
radicals that were in thermal equilibrium with the neutral gas and
only recently were excited by an electron impact or a collision with a
metastable particle (argon or nitrogen). 
The rotational distribution of this group should
be in good agreement with the translational temperature of the neutral
gas. The hot group consists mostly of radicals in the excited
electronic state (A) that appeared as an immediate result of water
dissociation. During this process, the water molecule bent at
ca. 104$^{\circ}$ gets excited to the B\,$^1A_1$ state, where the
angle straightens to 180$^{\circ}$. This is followed by
dissociation to rapidly rotating OH(A) radical and a hydrogen atom.

\begin{figure}
\begin{center}
  \includegraphics[width = 0.45\textwidth]{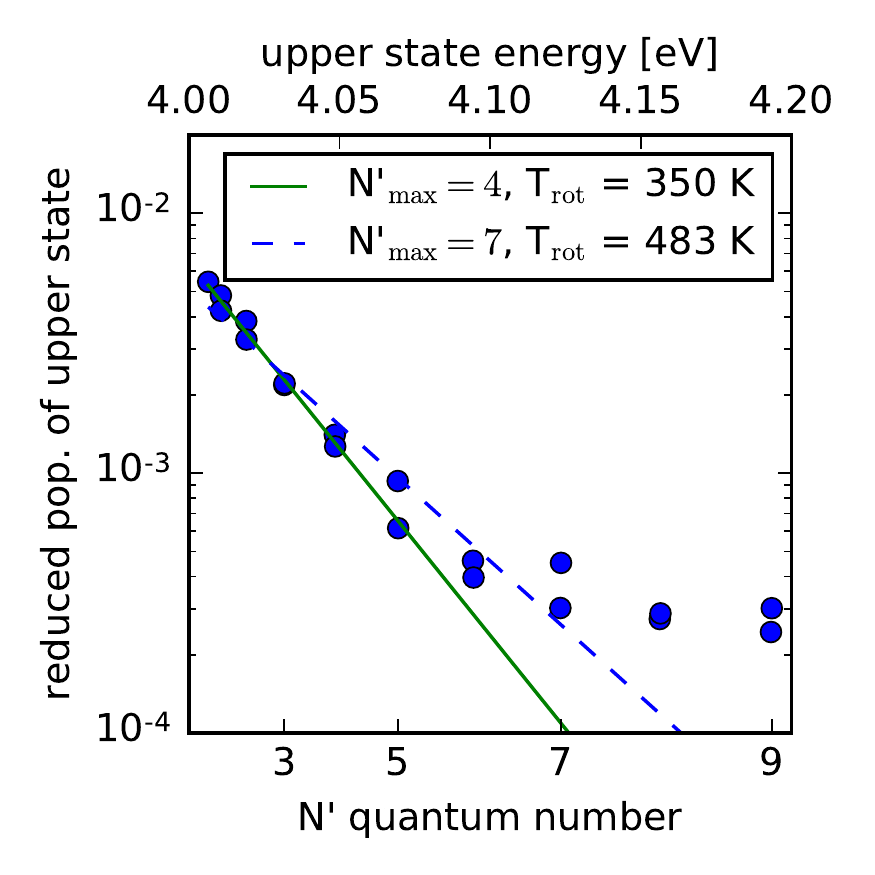}
  \end{center}
  \caption{Finding neutral gas temperature from OH Boltzmann plots by
    fitting a single-temperature Boltzmann distribution strongly
    depends on the arbitrarily chosen $N'_{\rm max}$.}
  \label{f:oneexp_bad}
\end{figure}

\begin{figure}
\begin{center}
  \includegraphics[width = 0.5\textwidth]{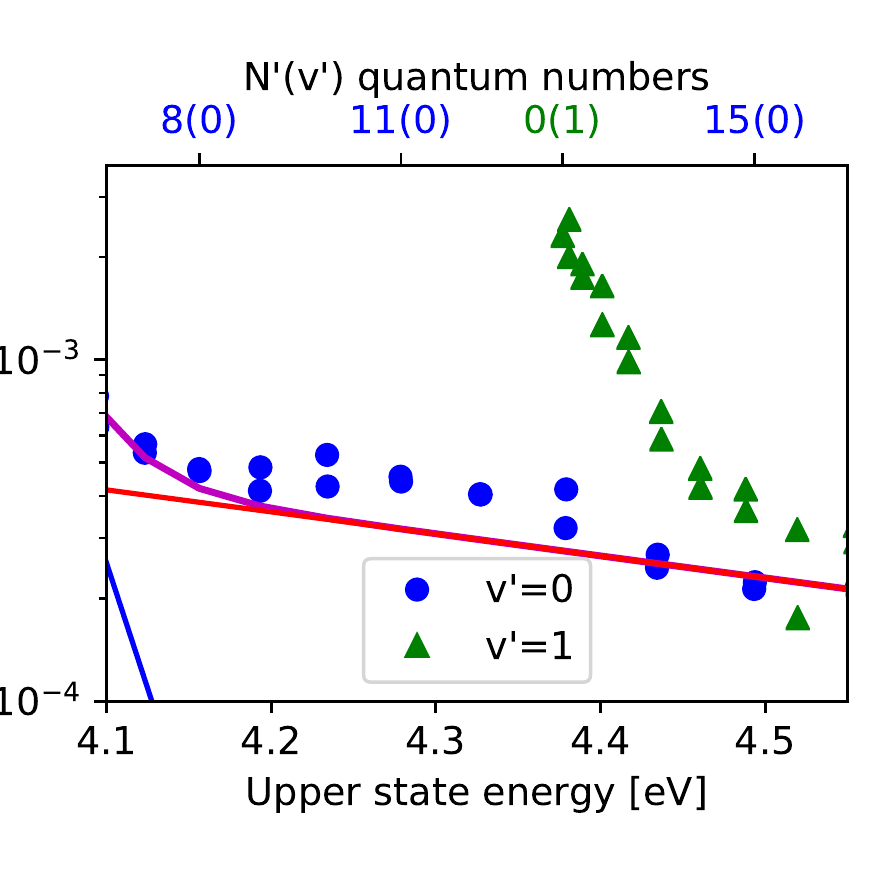}
  \end{center}
  \caption[]{ Detail of the Boltzmann plot from figure~\ref{f:OH_boltz}.
    The population of the states with $9 \leq N' \leq 13$
    deviates from the two-temperature
    distribution and they were excluded from the fit.}
  \label{f:OH_boltz_detail}
\end{figure}

In the literature, it seems that the traditional way is fitting a single exponential 
to states with $N'< N'_{\rm max}$~\cite{brug2,bruggeman2009rotational,nikiforov2011physical}, where $N_{\rm max}$
is typically around 10 (few units above or below, depending on the
discharge conditions). Such approach to our data gives results in the
range (350--480)\,K, mostly depending on the arbitrarily chosen value of
$N_{\rm max}$, see figure~\ref{f:oneexp_bad}. The temperature range
satisfactorily overlaps with the results from N$_2$(C) spectra,
compare with figure~\ref{f:N2T}. 
However, such approach is inconsistent with Ochkin's explanation of the rotational distribution. 
The observed population distribution should be a sum of (at least) two
independent ensembles of OH radicals in the A state. To reliably
obtain the rotational temperature of both, we have thus fitted the observed
population distribution with a sum of two independent Boltzmann
distributions. 
This is in principal similar to the treatment proposed by Bernatskiy~\cite{bernatskiy2015measurements}. 
A few implementation details, however, should be noted here.

The traditional way of first taking a logarithm of the reduced
populations and then fitting a straight line could not be used here. 
Indeed, a sum of two exponentials cannot be linearised by taking a
logarithm. Fitting a sum of exponentials to the distribution of the 
reduced populations, however, tends to fail, as the  populations of
states with higher rotational quantum numbers are orders of magnitude
smaller and their contribution to the total sum of squares is rather
negligible. 
%\textcolor{red}{dalsi veta neni jasna a ve tretim clenu rovnice je mala zavorka chudera sama, chyba? jak se dojde ke tretimu clenu?}
Even though taking a logarithm does not linearise the
problem, it can be used the same way as it is used in plots -- to
reduce the range of data and enhance the weight of small numbers. 
For this reason, the minimised expression was %thus 
%\begin{equation}
\begin{dmath}
%\begin{center}
  %\sum \limits_{J'} \left[n_{(J',N',v')}^{\rm theor} -
  % n_{(J',N',v')}^{\rm measured} \right]^2 = \\
 \min \limits _ {(a_1, a_2, T_1, T_2)}\sum \limits_{(J',N')} \left[  \ln \left( a_1\, 
  \e^{ -E_{(J',N')}/(k\,T_1)} +\\
  a_2\,\e^{-E_{(J',N')}/(k\,T_2)}\right)  - 
\ln \left(\frac {n_{(J',N')}}{2\,J'+1}   \right)  \right]^2,
%\end{equation}er
%\end{center}
\end{dmath}
$a_1$ and $a_2$ are the linear factors for the cold and hot group,
respectively, $T_1$  and $T_2$ are their respective rotational
temperatures. $E_{(J',N')}$ is the potential energy of each state in
question. 
This was performed for $v'=0$ only, so this quantum number is not
present in the state labels. It is not difficult to perform an
equivalent analysis also for higher vibrational states. For such
purpose, however, the raw spectra should be measured with greater
emphasis on the weaker lines. In this work, for the case of the surface streamer barrier discharge, 
the signal-to-noise ratio is sufficient only for the analysis of
$v'=0$.

Closer look at the population distribution around 4.3\,eV reveals a
third group of states, see figure~\ref{f:OH_boltz_detail}. 
The states with $v'=0$ and $9 \leq N' \leq
13$ are clearly overpopulated with respect to expectation from the
two-temperature fit. To increase the reliability of the
two-exponential fit, these states were excluded from the fitting
process.  
{
The analysis of OH(A-X) spectrum assuming multi-temperature
distributions was performed also by Bruggeman {\it et
    al.}, see~\cite{brug2}, figure~20, though the selected approach
  was not able to reveal any deviations from two-Boltzmann distributions.}
All the lines originating from these
states were carefully investigated using the Boltzmann-plot inspector
tool of the {\it massiveOES} software. The match of the fitted
distribution and the measurement at all spectral regions is strongly
convincing that this is a real observation and not an artefact of the
method. Possible causes of such overpopulation of this narrow
group of rotational states are yet to be found. The selection rules
for photon absorption, however, exclude an intense radiation at
ca. 4.3\,eV, as photons are not allowed to change the rotational
states by more than $J \pm 1$. It is more likely, that the residual
energy after collision-induced vibrational energy transfer ($v' = 1
\rightarrow 0$) is partially conserved in the rotational states. 
{In environments closer to thermal equilibrium, the
  population of ($v'=1$) states is usually orders of magnitude smaller
and such effect would not influence the population of $(v'=0)$ states
significantly.}
In
this case, however, the population of the $(v'=1)$ state is comparable to the
population of ($v'=0$) state and such effect could indeed be
observable. To our best knowledge, this is for the first time such 
effect has been observed -- thanks to the precise analysis enabled 
by the developed software. 
However, further investigations are desirable for detailed clarification.

\begin{figure}
\begin{center}
  \includegraphics[width = 0.5\textwidth]{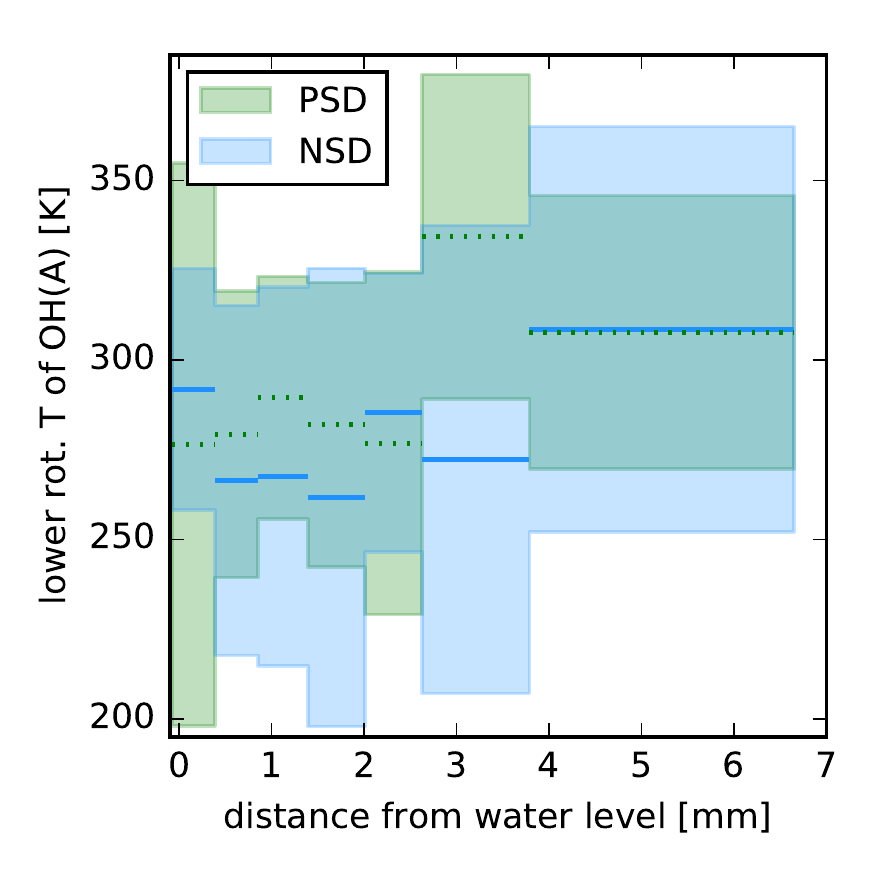}
  \end{center}
  \caption[]{Gas temperature from OH spectra, the lower rotational temperature. PSD (dotted) -- positive streamer discharges, NSD (solid) --
    negative streamer discharges.}
  \label{f:OHT}. 
\end{figure}

\begin{figure}
\begin{center}
  \includegraphics[width = 0.5\textwidth]{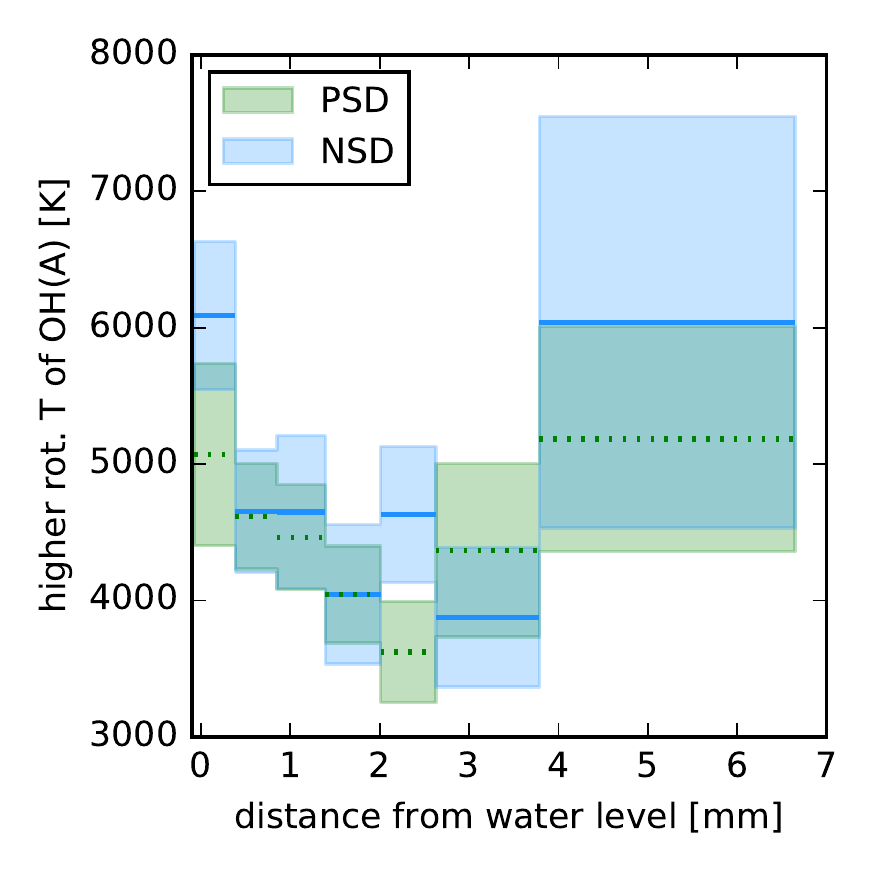}
  \end{center}
  \caption[]{The higher rotational temperature of OH(A). PSD (dotted) -- positive streamer discharges, NSD (solid) --
    negative streamer discharges.}
  \label{f:OHT_tail}. 
\end{figure}

The fitted temperatures from the Boltzmann plots are shown in
figures~\ref{f:OHT}~and~\ref{f:OHT_tail}. The lower rotational
temperature (figure~\ref{f:OHT}) should hold information about the temperature of the
neutral gas and it is expected to agree with the $T_{\rm rot}$ of
N$_2$(C) shown in figure~\ref{f:N2T}. Some characteristics are,
indeed, shared -- the temperature in PSD
appears to be slightly higher except for the triple-line interface,
the temperature profile is concave for NSD and approximately rising
for PSD. The absolute values of temperature from OH spectra are,
however, about 100 degrees lower. In some cases, the found temperature
is even below the expected ambient temperature. Based on the
observations presented in this article, it is not possible to
precisely determine the origin of this. The match of the
two-temperature fit to the measurement, shown in
figure~\ref{f:OH_boltz} supports the two-temperature
hypothesis. The rotational temperature of the {\it cold group} may differ from
the translational temperature of the gas. Although it is not quite
expected, it should not be excluded in this highly non-equilibrium
environment. 

The temperature of the {\it hot group} is shown in
figure~\ref{f:OHT_tail}. The interpretation of this parameter is not
quite straightforward. Nevertheless, we can compare it to the results
of other groups. Bruggeman~{\it et al.}~\cite{bruggeman2009rotational} 
have observed temperatures around 8000\,K, Ochkin reported
temperature of 9000\,K~\cite{ochkin2009spectroscopy}. 
Mohlmann~{\it et al.} have reported
temperature as high as 30\,000\,K -- in their experiment, they
dissociated water vapour by 100\,keV electron beam at low 
pressure~\cite{mohlmann1976rotational}.
 Vor\'a\v{c}~{\it et al.} have
observed a single-Boltzmann distribution in a microwave plasma
jet~\cite{vorac2017batch}. The rotational temperatures of
(2\,000--5\,500)\,K well matched with vibrational temperature of
OH(A). 

\begin{figure}
\begin{center}
  \includegraphics[width = 0.5\textwidth]{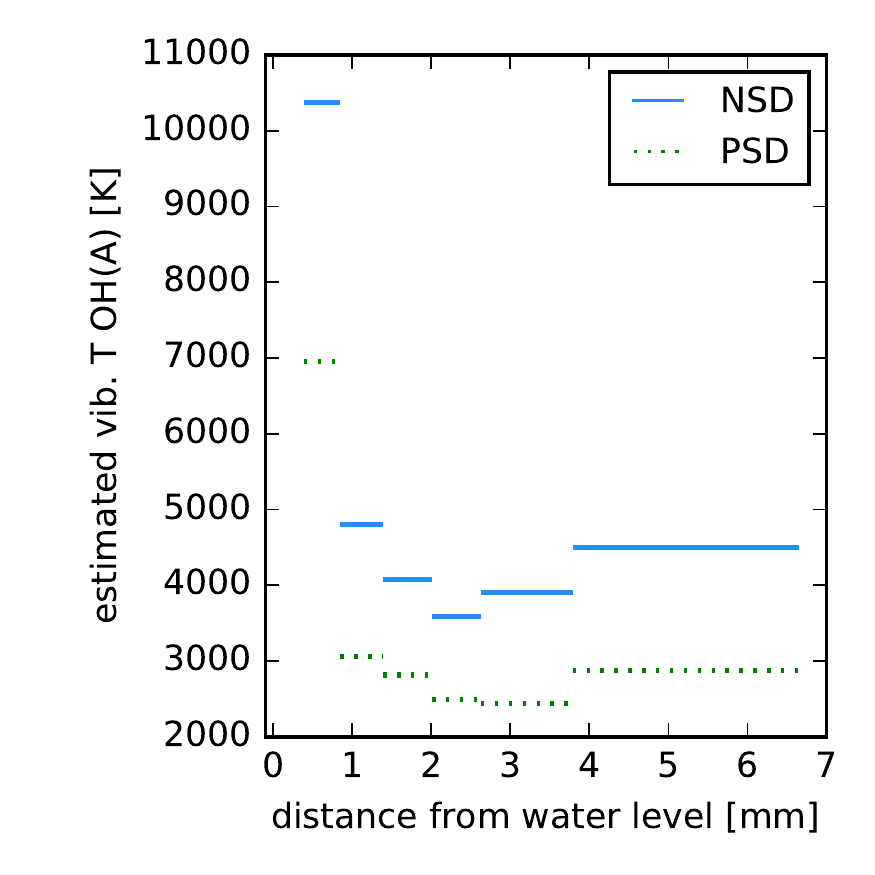}
  \end{center}
  \caption[]{The estimate of vibrational temperature of OH(A). PSD (dotted) -- positive streamer discharges, NSD (solid) --
    negative streamer discharges.}
  \label{f:OHTvib}. 
\end{figure}

In this work, we have estimated vibrational temperature
from the ratio of populations of $v'=0$ and $v'=1$ states, see
figure~\ref{f:OHTvib}. 
The intensity of lines from $v'=2$ is insufficient for such analysis. Near
the triple-line, the population of $v'=1$ state even exceeds the
population of $v'=0$ state and the ``vibrational temperature'' acquires
negative values for both polarities (not shown in~\ref{f:OHTvib}). The spatial profile qualitatively
well agrees with the spatial profile of tail (higher) rotational temperature --
both are concave and in both cases, the temperature of negative
streamer discharges seems to be higher. This is in contrary to the gas
temperature. This can be interpreted as follows. The electron-heavy
particles collisions are more likely to excite the electronic and
vibrational states of the molecules. Afterwards, the energy during
heavy particle collisions gradually changes to rotational and
translational form. 
At the time of light emission, 
 the relaxation from electronic and vibrational
states to rotations and translational motion (often called E-T, E-R,
V-T and V-R relaxation) has advanced further in the PSD than in the
NSD. The concentrations of collision partners are not expected to
change significantly during the discharge period. Therefore, this is
to be attributed rather to the duration of the discharges. This is
around 15\,ns for NSD and around 30\,ns for PSD, see
section~\ref{s:electro} and figure~\ref{el_detail}. 
In this respect the vibrational temperature and the tail rotational
temperature both seem to represent an intermediate step between the
direct electronic excitation and the final relaxation. They both can
be probably used as a vague lower limit for electron temperature.

\begin{figure}
\begin{center}
  \includegraphics[width = 0.5\textwidth]{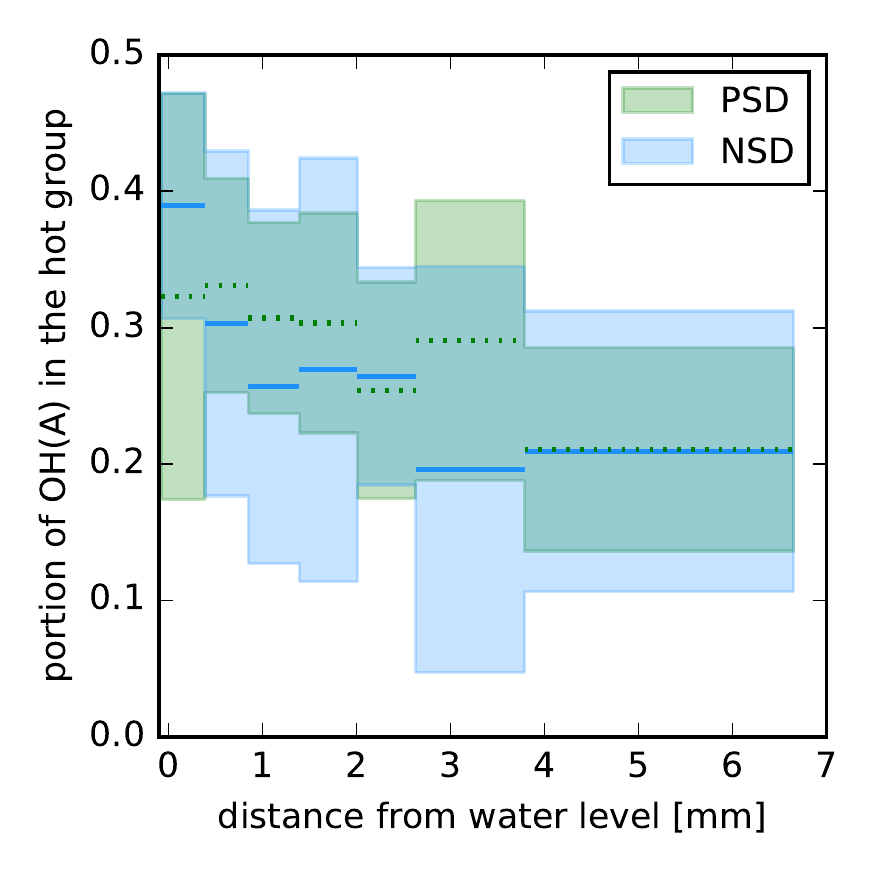}
  \end{center}
  \caption[]{Portion of OH(A) molecules in the hot group.
    PSD (dotted) -- positive streamer discharges, NSD (solid) --
    negative streamer discharges.}
  \label{f:hot_over_cold}. 
\end{figure}

Another interesting parameter is the portion of OH(A) molecules in the
hot group $\eta_{\rm hot}$ (figure~\ref{f:hot_over_cold}) calculated as
\begin{equation}
\eta_{\rm hot} = \frac{n_{\rm hot}}{n_{\rm hot} + n_{\rm cold}},
\end{equation} 
where $n_{\rm hot}$ and $n_{\rm cold}$ are the sums of the reduced
populations over all rotational states for the hot and cold group, respectively.
 $\eta_{\rm hot}$ reflects the OH excitation pathways. At the triple-line, over
one third of OH(A) molecules is found in the hot group. These should
be a direct product of water dissociation via the B\,$^1A_1$
state. This portion monotonously falls with the distance from the
water level to approximately one fifth at 5\,mm distance. It should be
noted, that water can be dissociated also other ways than via the B\,$^1A_1$
state~\cite{carr2014oh}. These are then expected to contribute rather to the cold
group and perhaps they may also lead to the observed low rotational
temperature of the cold group. In this respect, more investigations
are desirable. 
%To our best knowledge, this is the first time the
%rotational distribution was investigated in such thorough way. 

% \textcolor{red}{koukni Honzo do citace [17] na obr.20!!! delaji taky pro ruzny vibrace ruzny teploty...}

One of the aims of this work was to compare the estimates of neutral
gas temperature from rotational spectra of N$_2$(C-B) (figure~\ref{f:N2T})
and OH(A-X) (figure~\ref{f:OHT}).  There are several approaches to estimate this parameter
from the rotational spectra of OH(A-X) transition. The
single-Boltzmann spectra fitting (as performed here for N$_2$(C-B)) is
nowadays already considered unreliable, as was shown 
previously in~\cite{bruggeman2009rotational, nikiforov2011physical}
as well as in this work (figure~\ref{f:OH_fit}) -- the fitted
temperature of 848\,K does not reflect any meaningful physical
property of the discharge. It is also clear from the mismatch of the
single-Boltzmann simulation and the measurement. 
Preparing a Boltzmann plot and analysing it is thus
recommended. Fitting a single-Boltzmann distribution to few
arbitrarily selected states with low rotational quantum numbers may be
a misleading practice, as the value of result strongly depends on the
chosen states. If our understanding of the problem is correct and the
rotational distribution is really a sum of two distributions (cold and
hot group), the two-Boltzmann fit should be superior. In this case,
however, the temperature of the cold group does not agree with the
rotational temperature of N$_2$(C). This may show an unknown
process affecting the rotational distribution. In other words, the
resulting rotational distribution may be a sum of more than just two
distributions.

It should be also noted, that the shown errors in figures~\ref{f:N2T}
and~\ref{f:OHT} greatly differ in magnitude. The errors in the case of
N$_2$(C) are rather underestimated, see above. Furthermore, the number
of degrees of freedom in case of fitting a single-Boltzmann
distribution to a whole measured spectrum is much greater than in the
case of analysing a Boltzmann plot. This further suppresses the
calculated error for N$_2$(C) temperature. 
{The calculated error in the
case of temperatures from the OH Boltzmann plots, on the other hand, seems to be much
greater.} This, however, reflects the reality as the number of points
for the fit is much lower than in the case of the whole
spectrum. Also, the four fit parameters ($a_1$, $a_2$, $T_1$, $T_2$)
are rather strongly correlated -- i.e. the change of one can be
partially compensated by change of other. In this respect, N$_2$(C-B)
spectra seem as a more reliable way of determining the neutral gas
temperature. On the other hand, in cases where adding another specie
into the discharge is unacceptable, OH spectra may be the only way how to
learn something about the neutral gas temperature. In such cases, the
proposed way of two-Boltzmann fitting may yield plausible results.

\section{Summary and conclusion}
A special case of a surface barrier discharge in contact with a solid
dielectric and deionised water level was investigated. The driving
voltage was sinusoidal. Electrical investigations and ICCD imaging
have shown that the discharges appearing in the opposite half-cycles
are of fundamentally different nature. The discharges always start at the
triple-line interface in contact with the water level. At one
half-cycle, positive streamer discharges appear (average duration
30\,ns). The negative streamers (average duration 15\,ns) emerge in 
the other half-cycle. Optical emission spectroscopy was performed
for the respective half-cycles separately. Spectra of N$_2$(C-B) and
OH(A-X) transitions were recorded and analysed. N$_2$(C) appeared to
be in rotational equilibrium. Assuming, the rotational temperature
reflects the temperature of the neutral gas, a temperature profile
along the discharge path was constructed for the separate
half-cycles. It appeared that the positive streamer discharges are
slightly warmer, except for the triple-line interface, where the
temperature was comparable. The temperature in the discharges
was in the range (340--440)\,K. 

OH(A-X) spectra revealed a greatly non-equilibrial
rotational-vibrational distribution. A novel method of state-by-state
fitting has been developed and used to construct Boltzmann plots for
each measured spectrum. 
Preparing a Boltzmann plot may be a difficult task, particularly for overlapping spectra, as
noted also in~\cite{van2012temperature}. The proposed state-by-state
fitting offers a convenient way to construct Boltzmann plots even from
partially overlapping spectra. As this functionality is offered for
free as a part of an open-source software package {\it
  massiveOES}~\cite{massiveOES, vorac2017batch}, it is also quite
undemanding and available to the community. 

The Boltzmann plots were analysed assuming that the observed
distribution is a sum of two Boltzmann-distributed independent
populations. The states with $(v'=0,~9\leq N' \leq 13)$ were found to
be overpopulated with respect to the two-Boltzmann distribution,
perhaps due to fast vibrational-energy transfer.  The novel approach
to the OH(A-X) spectra enabled us to decouple the different groups of
OH(A) and thoroughly investigate their properties. The temperature of
the hot group of OH(A) was linked with the vibrational temperature
estimate, as they both are believed to show an intermediate step of
the E-T and E-R relaxation. This is further supported by the
observation, that in the longer lasting positive streamer discharges,
both vibrational and hot group rotational temperatures were lower
(compared to NSD), while the gas temperature was higher.

The OH(A-X) spectrum was found to be able to provide a rough estimate of
the neutral gas temperature. 
However, due to presence of several indistinguishable excitation
pathways that severely influenced the resulting rotational
distribution, it is not the preferable way. 
Whenever possible, thermometry using molecules that are less influenced by
chemistry is recommended.
Nevertheless, thorough investigations of OH(A-X) spectra with enhanced
temporal resolution may bring important insights into the plasma
process in transient discharges.

\section*{Acknowledgements}

This research was funded by the Czech Science Foundation project
16-09721Y.  This research has been also supported by the 
project LO1411 (NPU I) funded by Ministry of Education Youth and
Sports of Czech Republic.

The graphs in this publication were composed in the Matplotlib 2D graphics
environment \cite{hunter2007matplotlib}.

\appendix
\section{Comment on state-by-state fitting}
The simulated fractional spectra $s_{J'}(\lambda, G, L)$ are functions
of wavelength $\lambda$, and the Gaussian ($G$) and Lorentzian ($L$) line
broadening parameters. They are thus constant for a particular 
spectrometer settings (unless a line broadening mechanism comparable
to or stronger than the instrumental broadening is present). 

The linearity of the problem greatly enhances the computational
feasibility and a usual office computer can perform the task in a few
seconds. \verb|scipy.optimize.nnls()| function, an implementation of
algorithm described in~\cite{lawson1995solving}, is used for the fit.
This approach also allows constructing Boltzmann-plots from spectra with
moderate resolution with partially overlapping
lines. These are notoriously problematic to obtain from the height 
of the line~\cite{van2012temperature}. 

Among the known limitations is particularly the requirement
that the number of the considered states must not exceed the number of
pixels (the number of the degrees of freedom must be positive). It is
also advisable to analyse spectra of molecules with not too few
observable lines from one state, as this improves the reliability of
the result. A typical example of spectrum not appropriate for this
procedure are the bands of N$_2$(C-B) transition. These bands have
strongly suppressed Q-branch and almost completely overlapping
P-branch. Effectively, only one line in the R-branch
is available for each state, making the fit vulnerable to noise or
overlapping lines. Precise wavelength matching of the simulation and
the measurement is, of course, essential.

\section*{References}
\bibliography{literature}{}
\bibliographystyle{iopart-num}

\end{document}